\documentclass[12pt]{article}
\usepackage{graphicx}
\begin{document}

\begin{center}
{\bf \Large Gaseous Detector of Ionizing Eradiation in Search for
Coherent Neutrino-Nucleus Scattering}
\vskip 0.2in

A.V.Kopylov, I.V.Orekhov, V.V.Petukhov, A.E.Solomatin \\
{\it Institute of Nuclear Research of Russian Academy of Sciences\\
117312 Moscow, Prospect of 60th Anniversary of October Revolution
7A}
\vskip 0.2in

\end{center}

\footnotetext{Corresponding author: Kopylov A.V., Institute for
Nuclear Research of Russian Academy of Sciences, Prospect of 60th
Anniversary of October Revolution 7A, 117312, Moscow, Russia;
telephone +7(495)8510961, e-mail: beril@inr.ru }

\begin{abstract}
We propose to search for coherent neutrino-nucleus scattering (CNNS)
by means of a triple-sectioned low background proportional counter.
As a working medium we plan to use argon and xenon at about 1 MPa.
We have shown using bench-scale assembly, that pulse-shape
discrimination enables to effectively suppress noise pulses from
electromagnetic disturbances and microphonic effect in the energy
region where one expects signal from CNNS (from 20 eV to 100 eV)
with a factor of about $10^3$. The calculation has been done of the
background from neutrons, generated by muons of cosmic rays. The
experimental setup has been proposed.
\end{abstract}

At small recoil energies when neutrino does not ``see'' nucleons
constituting a nucleus but rather scatters as a wave on a grid, the
neutrino-nucleus elastic scattering by means of exchange of
$Z_0$-bozon is coherent over the nucleons in the nucleus. Due to the
coherence the cross section is proportional to the square of the
number of neutrons in a given nucleus (the contribution of protons
is given in the expression for the cross section with the weight of
approximately 0.08) and it can reach so big value that even for a
mass of a target of about 1 kg and a flux of antineutrinos from a
reactor of $2\cdot10^{13} nu/cm^2/s$ the count rate may reach the
value of several events per day. This process has been described in
70th of last century \cite{1,2}, has been often discussed later on
\cite{3}--\cite{7} but has never been observed because of extremely
small (less than 600 eV for reactor antineutrinos) kinetic energy of
the recoiling nucleus and only small portion of this energy (about
15\%) is transferred into the one of ionizing eradiation. The
discovery of this process would be the great achievement of the
modern physics and this explains the current interest of
experimentalists to this task.  By choosing the gaseous proportional
counter as a detector of CNNS the emphasis is done on the following
advantages of this technique:

\begin{enumerate}
\item  Very high factor of the gas amplification ($> 10^4$).
\item  Possibility to use gas at relatively high pressure about 1 MPa to
obtain the mass sufficient for count rate of about 1 events per day.
\item  Good signature of the events by a pulse shape (very characteristic
front and tail of the pulses).
\item  The possibility to discriminate noise from electromagnetic
disturbances and microphonic effect.
\item  Availability of the efficient methods of gas purification.
\item  Detector can be fabricated only from very pure materials without
PMTs as a possible source of ionizing eradiation etc.
\item  The possibility easily change the working gas (argon -- xenon)
not changing the configuration, what is important to perform the
comparative measurements at the same site.
\end{enumerate}

We performed the measurements of the energy spectra of the pulses in
argon using a small bench scale assembly. The calibration has been
done using $^{55}Fe$ as a source of X-ray eradiation of 5.9 keV.
Proportional counter had 37 mm, the central wire of 20 mm in
diameter and it was filled by argon and methane (10\%) mixture by
100 and 300 kPa. The shapes of the pulses from output of charge
sensitive preamplifier of the sensitivity of about 0.4 V/pC have
been recorded by 8-bit digitizer. The shapes recorded during certain
time were analysed in off-line. The aim was to see how efficient
could be the pulse shape discrimination of the noise pulses from
electromagnetic disturbances and microphonic effect in the region
below 100 eV, i.e. where the main effect is expected from CNNS of
reactor antineutrinos. In Fig.1 we show the pulses observed during
time interval $400 \mu s$ where one can see ``true'' pulse with
correct signature from ionization process and ``wrong'' pulse from
electromagnetic disturbances.

\begin{figure}[!ht]
\centering
\includegraphics[width=4in]{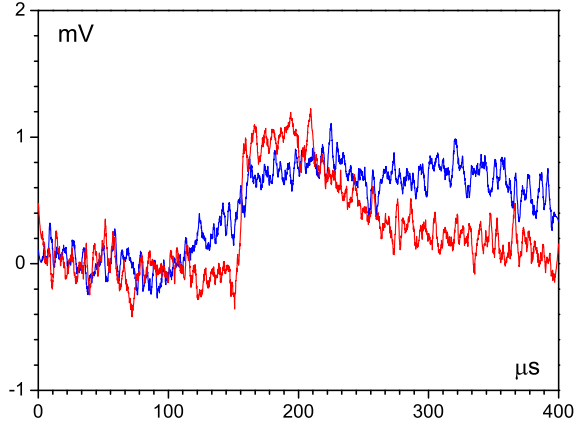}
\caption{The pulses on the output of charge sensitive preamplifier:
The point ionization (red) and from electromagnetic
disturbances(blue). The sensitivity of electronic channel 1 mV/10
eV}
\end{figure}

Electromagnetic disturbances have usually non regular shape, the
pulses from ``microphonic effect'' have typically response in the
audible range with the shapes close to sinusoidal. The pulses from
the point ionization in gas have typically a relatively short front
edge (a few microseconds) corresponding to the time drift of
positive ions to cathode and long (hundreds of microseconds) tail
corresponding to the time of the base line restoration of the charge
sensitive preamplifier. These events might be produced in our
detector by internal radioactivity of the materials of the counter,
by electronic emission from the walls of the counter and also by
ionizing particles produced by cosmic rays. The amplitude of these
events may be even smaller then an average energy to produce a
single electron pulse because of the relatively broad energy
distribution in this case (Polia distribution). Using two peaks from
$^{55}Fe$ calibration source (5.9 keV and 2.85 keV escape peak in
argon) we observed relatively good linearity of the conversion
energy -- amplitude and rather high $4\cdot10^4$ gas amplification.
In the range from 20 eV to 100 eV, where main effect from coherent
scattering of reactor antineutrinos should be observed, the pulse
shape discrimination enabled to reduce the noise by a factor of
about $10^3$. Thus we show that this range can be effectively used
for counting of the events from CNNS. The similar problem of
counting the events from very small energy release has been solved
in a number of experiments with cryogenic detectors. In 1997 we
together with the staff of the laboratory of  Professor Sandro
Vitale in University of Genoa in Italy were first who succeeded in
counting the pulses from peaks 57 eV and 112 eV from the decay of
$^7Be$ \cite{8}. The energy threshold in this work was 40 eV. This
was achieved thanks to effective pulse shape discrimination of the
noise pulses from electromagnetic disturbances and ``microphonic
effect''. The count rate from CNNS of reactor antineutrinos  is
calculated to be a few events per day per kg of argon in the energy
range from 20 eV to 100 eV. To collect a mass of argon of about 1 kg
the detector should have the volume of about 50 liters even at the
pressure 1 MPa. But to get the gas amplification higher $10^4$ at
High Voltage 3 kV the diameter of the cathode should be 40 mm, not
more. To reconcile these conflicting demands we should use an array
of counters and each counter should have a central, avalanche region
with a small diameter of the cathode and external, drift region,
separated from avalanche region by a grid. The diameter of the drift
region is taken to be 140 mm. Apart from this, there should be
external cylindrical layer of counters working as an active
shielding and also as a passive one of the fluorescence from the
walls of the counter. All assembly is placed in a cylindrical body
made of titanium as a relatively pure on $^{226}Ra$ material, as our
previous measurements have shown. In Fig.2 we show the general view
of this counter.

\begin{figure}[!ht]
\centering
\includegraphics[width=4in]{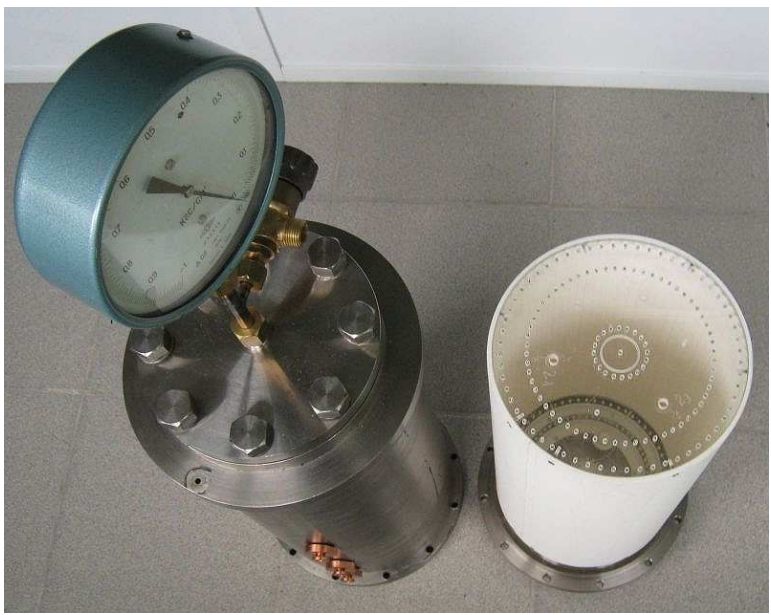}
\caption{The detector before assembling}
\end{figure}

We plan to use an array of 16 similar counters, each working on
separate charge sensitive preamplifier and digitizing board. The
counters will be assembled in 4 planes, each one having 4 counters.
The size of the assembly will be approximately 100x100x100 cm. To
reduce the background from cosmic rays, neutrons and gamma-rays the
assembly will be placed in the box made of slabs of iron 30 cm
thick, internal surfaces will be lined by borated polyethylene 20 cm
thick. To shield from fast neutrons from the reactor we plan to use
additional external layer of water 50 cm thick and on the outside --
plastic scintillator as an active veto shield from ionizing
particles of cosmic rays penetrating to the depth of about 16 m of
water equivalent. The water shield reduces the background from fast
neutrons by an order of magnitude, thus, it will be possible by
comparing the data collected with and without water to determine how
large will be the contribution of reactor neutrons to the effect
observed. All this assembly will be placed in a hermetically sealed
housing filled by argon purified of radon. We select this design of
shielding to reduce at most the background from gamma-quanta from
external radioactivity and from neutrons, generated in iron by
cosmic rays. Borated polyethylene 20 cm thick decreases
approximately 10-fold the flux of fast neutrons from iron. The slabs
of iron 30 cm thick effectively absorb gamma-radiation from the
walls. In Fig.3 we show the calculated effect from CNNS and the
background from neutrons, generated at 16 meters of water equivalent
for argon and xenon as a working medium of the detector.

\begin{figure}[!ht]
\centering
\includegraphics[width=4in]{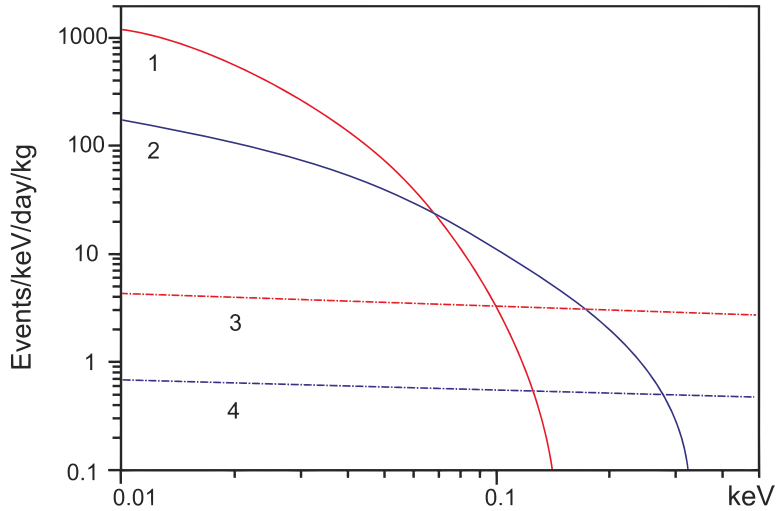}
\caption{The energy spectrum of nuclear recoils from CNNS of reactor
antineutrinos (1 -- xenon, 2 -- argon) and from scattering of
neutrons, generated by muons of cosmic rays (3 -- xenon, 4 --
argon)}
\end{figure}

The energy spectrum of nuclear recoils presented on Fig.3 was taken
from \cite{9}. In the calculation of the background from scattering
of neutrons, generated by muons of cosmic rays, we used the data on
the energy spectrum of neutrons  from \cite{10}. For precise
interpretation of the effect from CNNS one needs an accurate, with
the uncertainty of a few percents, measurement of the quenching
factor in gaseous xenon and argon which is expected to be
approximately 10-15\% at the energy of the recoiling nucleus lower
then 500 eV \cite{11}. Further development of the technique
described in this paper is needed to accomplish the task within
approximately 5 years to obtain some significant physical result.

\vskip 0.2in

{\bf Acknowledgements.} We warmly acknowledge funding from the
Programs of support of leading schools of Russia (grant
\#871.2012.2)


\begin{thebibliography}{99}

\bibitem{1} D.Z.Freedman PRD 1974 {\bf 9} 1389

\bibitem{2} D.Z.Freedman, D.N.Schramm, and D.L.Tubbs Ann.Rev.Part.Sci. 1977
{\bf 27} 167

\bibitem{3} A.Drukier and L.Stodolsky Phys.Rev.D 1984 {\bf 33} 2295

\bibitem{4} L.M.Krauss Physics Letters 1991 {\bf B269} 407

\bibitem{5} C.J.Horowitz, K.J.Coakley, D.N.McKinsey Phys.Rev. 2003
{\bf D68} 023005

\bibitem{6} K.Patton, J.Engel, G.C.McLaughlin, and N.Schunk 2012
Phys.Rev. {\bf C86} 024612

\bibitem{7} V.F.Shvartsman, V.B.Braginskii, S.S.Gershtein, Ya.B.Zeldovich and M.Yu.Khlopov
JETP Lett. 1982 {\bf 36} 277

\bibitem{8} M.Galeazzi,...  A.V.Kopylov, V.V.Petukhov et al. Phys.Lett.
1998 {\bf B398} 187

\bibitem{9} D.Akimov, A.Bondar, A.Burenkov and A.Buzulutskov 2009
[arXiv:0903.4821]

\bibitem{10} T.Perera {\it PhD thesis, Department of Physics} Case Western Reserve University 2002

\bibitem{11} K.W.Jones and H.W.Kraner Phys.Rev. 1975 {\bf A11} 1347

\end{thebibliography}
\end{document}